\documentclass[twocolumn, longbibliography, prl, aps, amsmath, amssymb,superscriptaddress]{revtex4-2}
\usepackage{amsmath}
\usepackage{amsfonts}
\usepackage{graphicx}
\usepackage[colorlinks,linkcolor=blue,citecolor=blue,urlcolor=blue]{hyperref}
\usepackage{float}
\usepackage[normalem]{ulem}
\usepackage[switch,columnwise]{lineno}
\usepackage[dvipsnames]{xcolor}

\begin{document}

\title{Arcsine Laws of Light}

\author{V. G. Ramesh}
\author{K. J. H. Peters}
\author{S. R. K. Rodriguez}
\affiliation{Center for Nanophotonics, AMOLF, Science Park 104, 1098 XG Amsterdam, Netherlands}

\date{\today}

\begin{abstract}
We demonstrate that light in a coherently driven resonator obeys L\'evy's arcsine laws --- a cornerstone of extreme value statistics. This behavior emerges asymptotically in the time-integrated transmitted intensity, an important quantity which is measured by every photodetector. We furthermore demonstrate a universal algebraic convergence to the arcsine laws as the integration time increases, independent of the balance between conservative and non-conservative forces exerted on the light field. Through numerical simulations we verify that the arcsine laws are also obeyed by the light field quadratures, and in a Kerr nonlinear resonator supporting non-Gaussian states of light. Our results are relevant to fundamental studies and technological applications of coherently driven resonators  (in e.g., optics, microwave photonics, and acoustics), which in turn open up perspectives for probing emergent statistical structure in new regimes and in systems with memory.
\end{abstract}

\maketitle

Random processes have fascinated physicists for decades. A seminal result due to Paul L\'evy is the existence of arcsine laws for random walks and Brownian motion~\cite{Levy40}. There are three arcsine laws, one for each of these observables: i) the fraction of time spent above the mean, $T_1$; ii) the fraction of time elapsed since last crossing the mean, $T_2$; and iii) the fraction of time taken to reach the maximum, $T_3$. For all three times $T_{j=1,2,3}$, the probability distribution $\mathcal{P}$  and cumulative distribution $\mathcal{C}$ are:
\begin{subequations}
\begin{align}
    \mathcal{P}(T_j) &= \frac{1}{\pi} \frac{1}{\sqrt{T_j(1-T_j)}} \label{eq1a}\\
    \mathcal{C}(T_j) &= \int_0^{T_j} \mathcal{P}(T_j') d T_j' = \frac{2}{\pi} \arcsin{\left(\sqrt{T_j}\right)}.
    \label{eq1b}
\end{align}
\end{subequations}
Equation~\ref{eq1b} inspired the name ``arcsine laws'', while Eq.~\ref{eq1a} expresses the interesting fact that extreme deviations from the mean are very likely to occur.

A wide variety of physical systems sustain  stochastic dynamics obeying the arcsine laws~\cite{Godreche01,Campos2003,Rebenshtok08,Oshanin09, Barato18,Mori19,Spiechowicz21,Singh22} . Also in competitive sports~\cite{Clauset15}, genomics~\cite{Fang21}, and finance~\cite{Han21,Dale80,Guo14}, the arcsine laws are obeyed. Generalizations of the arcsine laws for fractional Brownian motion have also been achieved~\cite{Sadhu18}. The form of the distribution~\ref{eq1a} has also drawn great interest in studies of wave transport through disordered media~\cite{Nazarov94, Jalabert94, Beenakker97, QCS, RotterGigan}, since it corresponds to the distribution of transmission eigenvalues. In this correspondence, the so-called open and closed channels (which dominate transport~\cite{Pendry90}) are the counterparts of the extreme deviations from the mean observed in $T_{j=1,2,3}$.

Despite their pervasiveness, the arcsine laws have never been explored in coherently driven resonators. Such resonators are an ideal platform for investigating extreme value statistics and the arcsine laws in hithero unexplored regimes. In particular, we will show that the balance between conservative and non-conservative forces acting on light fields can be precisely controlled across an enormous range. This control enables probing  emergent statistical structure in systems driven out-of and into equilibrium in arbitrary ways. In turn, a greater understanding of the statistics of noisy light fields can inspire new ideas and advances in technologies where optical noise plays a crucial role, like sensors~\cite{Langbein18, Lau18, Mortensen18, Rodriguez20, Duggan22}, beyond-von-Neumman computers~\cite{Kalinin18, Liew19, Opala19}, isolators~\cite{Shi15, Xuereb18, Rodriguez19, Yang20, Cotrufo21_1, Cotrufo21_2}, and quantum devices~\cite{Walmsley15, Blais18, Bai19, Hamerly20}.

In this Letter we demonstrate that the time-integrated transmission of a coherently driven optical resonator obeys the arcsine laws. Our experiments evidence that the arcsine laws hold for arbitrary driving conditions. These include protocols with a time-dependent non-conservative force which drives the system out of equilibrium.  Beyond demonstrating the arcsine laws asymptotically, we analyze how the finite time cumulative distributions of $T_{j=1,2,3}$ converge to $\mathcal{C}(T_{j=1,2,3})$  as the integration time increases. This convergence follows a power law with universal exponent independent of the system parameters and the extent of non-equilibrium behavior. Our experiments evidence the emergence of arcsine laws when $T_{j=1,2,3}$ are determined from the time-integrated intensity in a linear cavity; this is the quantity of interest in most optical experiments. In addition, through numerical simulations we verify that the arcsine laws also hold  for the time-integrated field quadratures (measurable via balanced homodyne detection and relevant to optimal sensing~\cite{Lau18}),  and in a Kerr-nonlinear resonator supporting non-Gaussian states of light.

\begin{figure}
    \centering
    \includegraphics{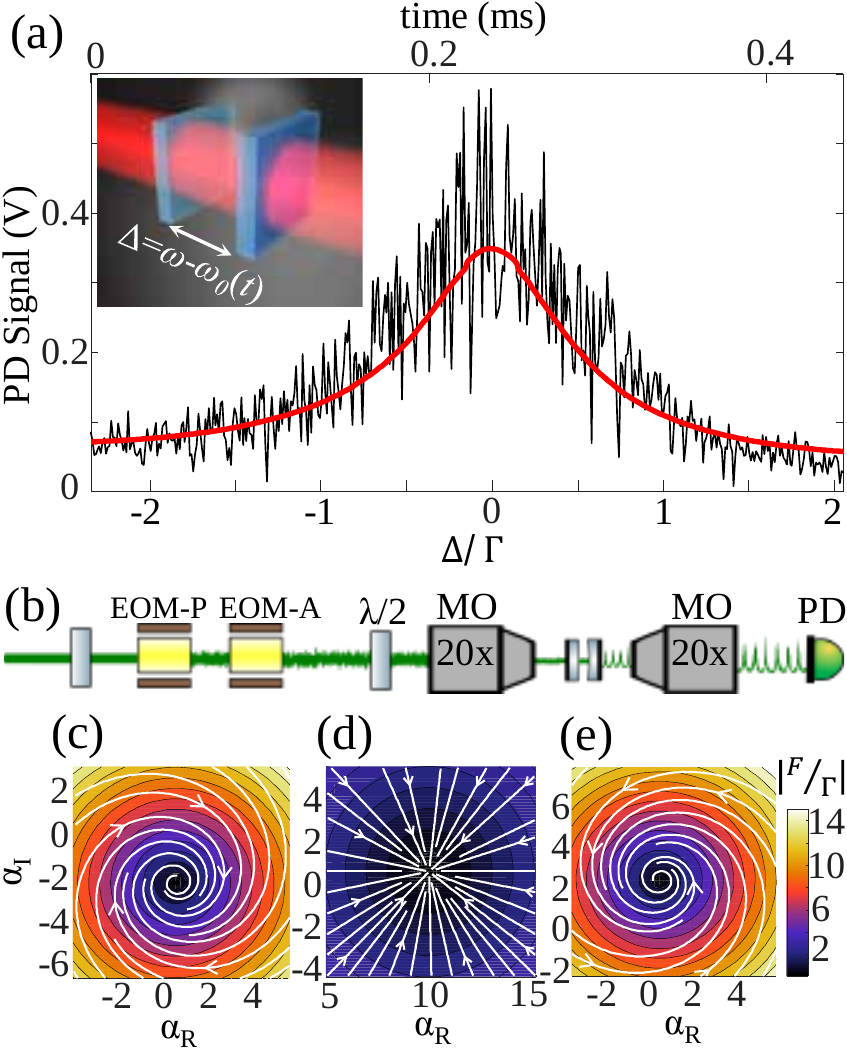}
    \caption{\label{fig1} (a) Inset: Optical system for experimentally testing the arcsine laws, namely a single-mode cavity driven by a continuous wave laser.  The cavity length and hence the laser-cavity detuning $\Delta$ are periodically modulated as a triangular wave. Main panel: Single-shot and averaged intensity transmitted through the cavity as black and red curve, respectively, both as a function of $\Delta$ referenced to the loss rate $\Gamma$. (b) Experimental setup for measuring the cavity transmission while adding noise to the laser amplitude and phase using electro-optic modulators EOM-A and EOM-P, respectively. MO means microscope objective and PD means photodetector.  (c)-(e) show the phase portrait of the system with color and white arrows representing the force magnitude $|F/\Gamma|$ and direction, respectively. $A/\sqrt{\Gamma}=7$ for all calculations, and (c) $\Delta/\Gamma=-2$, (d) $\Delta/\Gamma=0$, and (e) $\Delta/\Gamma=2$}
\end{figure}

Our experiment involves measuring the transmission of a continuous wave laser through a plano-concave Fabry-P\'erot cavity [see Fig.~\ref{fig1}(a) inset], while periodically modulating the cavity length. We use piezoelectric actuators to align the cavity mirrors and then to modulate their distance. The planar mirror is a 50 nm thick silver layer on glass. The concave mirror (5.2 $\mu$m diameter, 6 $\mu$m radius of curvature)  was made by focused ion beam milling a glass substrate~\cite{Trichet15}, and subsequently depositing a distributed Bragg reflector with $99.9$\% reflectance at the laser wavelength $532$ nm. The small radius of curvature and high mirror reflectivity strongly confine the optical modes. This allows us to probe a single optical mode in scans of up  to $\sim 50$ resonance linewidths; we implement shorter scans of 4.5 linewidths. In this single mode limit, the light field $\alpha$ in our cavity satisfies
\begin{equation} \label{eq2}
    i\dot\alpha = \left(-\Delta(t)- i\frac{\Gamma}{2}\right)\alpha + i\sqrt{\kappa_L}A + D\zeta(t).
\end{equation}
Equation~\ref{eq2} is written in a frame rotating at the laser frequency $\omega$. $\Delta(t)=\omega-\omega_0(t)$ is the detuning between $\omega$  and the resonance frequency $\omega_0(t)$, which we modulate using the actuators. $\Gamma = \gamma_a +\kappa_L+\kappa_R$ is the total loss rate, with $\gamma_a$ the absorption rate and $\kappa_{L,R}$ input-output rates through the left or right mirror. $A$ is the laser amplitude, assumed to be real. $\zeta(t)$ is a complex-valued Gaussian process representing white noise in the laser amplitude and phase; $D$  is its standard deviation.  Defining $\zeta(t)=\zeta_R(t) + i\zeta_I(t)$, the noise quadratures have mean $\langle\zeta_R(t)\rangle =\langle\zeta_I(t)\rangle=0$ and correlation $\langle\zeta_j(t')\zeta_k(t)\rangle = \delta_{j,k} \delta(t'-t)$.
For our linear coherently driven system, classical and quantum descriptions are statistically equivalent~\cite{Sudarshan63}. Hence,
Equation~\ref{eq2} remains a valid description of our cavity at arbitrarily low intra-cavity intensities.

Figure~\ref{fig1}(b) illustrates our setup, enabling fine control over every parameter in Eq.~\ref{eq2}.  We use microscope objectives with $20\times$ magnification and $0.4$ numerical aperture for light injection and transmission collection. The laser power entering the excitation objective is $1.25$ mW. This power is sufficiently high to minimize the effects of detector noise, yet sufficiently low to avoid nonlinearities. The  excitation path contains an amplitude and a phase modulator, each driven by a distinct waveform generator (not shown) to imprint noise on the laser. As shown in supplemental material, the power spectrum of the noise is flat across several decades~\cite{supp}. We implemented a 0.45 ms modulation period for $\Delta$ [see upper axis of Fig.~\ref{fig1}(a)] to operate within this flat range, but slightly closer to the low-frequency end. Consequently, our measurements are effectively influenced by white noise as in the model.

By modulating $\Delta$ we drastically change the force exerted on $\alpha$. This can be recognized by writing $\alpha=\alpha_R + i \alpha_I$ and decomposing Eq.~\ref{eq2} into real and imaginary parts:

\begin{align}
  \begin{pmatrix}\Dot{\alpha_R} \\ \Dot{\alpha_I}\end{pmatrix} =
      \underbrace{   \begin{pmatrix}
   -\frac{\Gamma}{2} & -\Delta \\
     \Delta &-\frac{\Gamma}{2}
   \end{pmatrix}
   \begin{pmatrix}
    \alpha_R\\ \alpha_I \end{pmatrix}  + \begin{pmatrix}
    \sqrt{\kappa_L}A\\ 0 \end{pmatrix}}_{F/\Gamma}+ D\begin{pmatrix}
    \zeta_R (t)\\ \zeta_I (t) \end{pmatrix}.
\label{eq: 2}
\end{align}

\noindent Equation~\ref{eq: 2} describes two overdamped anti-symmetrically coupled Langevin oscillators. The underbraced term is the deterministic force $F$ acting on the oscillators. It is divided by $\Gamma$ to recover the normal form of the overdamped Langevin equation. For $\Delta = 0$, $\alpha_R$ and $\alpha_I$ decouple and $F$ is conservative: $F$ can be derived from a scalar potential. Reference~\onlinecite{Busink} shows that, in this case, $\alpha_{R,I}$ each follow the Boltzmann distribution characterizing equilibrium behavior. In contrast, $\Delta \neq 0$ makes $F$  non-conservative:  $F$ cannot be derived from a scalar potential.

To illustrate the conservative vs. non-conservative character of the force on $\alpha$, Figs.~\ref{fig1}(c)-(e) show the phase portrait of the cavity at three instants of our protocol. The magnitude and direction of the force $F/\Gamma$ are encoded in color and white arrows, respectively. Figures~\ref{fig1}(c) and ~\ref{fig1}(e) correspond to values of $\Delta/\Gamma$ near the ends of our protocol, where the laser is far detuned from the cavity. The spiraling orbits, clockwise in Fig.~\ref{fig1}(c) and anti-clockwise in Fig.~\ref{fig1}(e), are typical of a non-conservative force. In contrast, in Fig.~\ref{fig1}(d) all force vectors are perpendicular to the contours of constant $|F/\Gamma|$ and point directly to the fixed point; this is the hallmark of gradient flow and equilibrium behavior.

Langevin dynamics in systems with non-conservative forces and anti-symmetric couplings have drawn significant attention in stochastic thermodynamics~\cite{Chernyak06,Blickle07,Borlenghi_2017}. However, tests of the arcsine laws in such systems have not been reported. Our experiment therefore tests the arcsine laws in a hitherto unexplored regime where the steady state transitions back and forth between equilibrium and non-equilibrium as $\Delta$ varies. The non-equilibrium behavior, due to the non-conservative force, is independent of the speed of the protocol. Actually, since the system's relaxation time is $\Gamma^{-1} \sim$ 1 ps and the scanning time is  $\sim 2.5$ ms, our entire protocol is adiabatic.

Figure~\ref{fig1}(a) shows measurements of the transmitted intensity when scanning the cavity length. The scan starts and ends where the non-conservative force dominates, as Figs.~\ref{fig1}(c,e) show. The black curve is the single-shot intensity. The red curve is the intensity averaged over 200 cycles, evidencing the Lorentzian lineshape expected for a linear resonator. Fluctuations in the single-shot intensity are due to the noise imprinted on the laser by the modulators. Compared to the intrinsic laser noise, the imprinted noise has a standard deviation that is $\sim 80$ times larger. In this way, we ensure that the dominant noise is the one we have characterized.

We are interested in the transmitted intensity integrated over $n$ modulation cycles, $\mathcal{I}_{n\tau} = \int_{0}^{n\tau} \kappa_R |\alpha(t)|^2 dt$ with $\tau$ the period.  We hypothesized that arcsine laws emerge in distributions of $\mathcal{I}_{n\tau}$ as $n \rightarrow \infty$. Our hypothesis was inspired by the work of Barato \emph{et al.}, where thermodynamic currents (e.g., the time-integrated work) were shown to obey the arcsine laws~\cite{Barato18}. However, unlike the experiment reported in Ref.~\onlinecite{Barato18} where a Brownian particle experiences a purely conservative force, the light field in our cavity undergoes both conservative and non-conservative dynamics within a single cycle.

\begin{figure}
    \centering
    \includegraphics{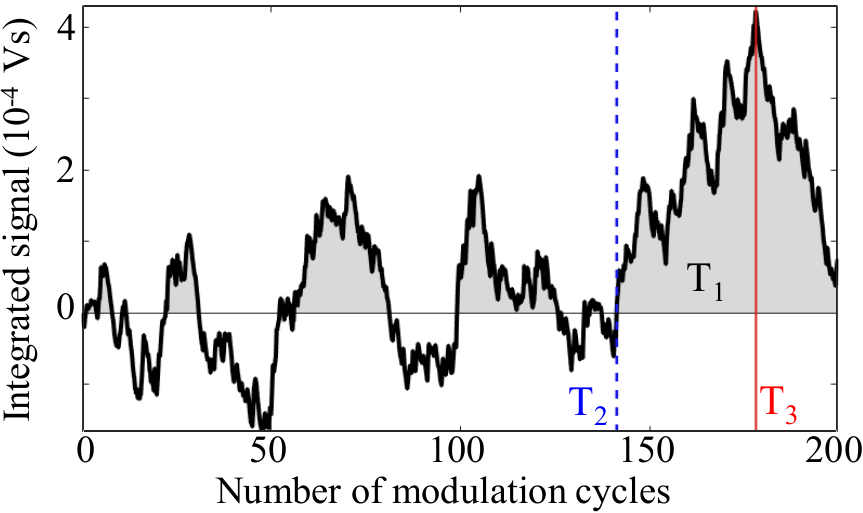}
    \caption{\label{fig2} Sample trajectory of the time-integrated transmitted intensity relative to its instantaneous average. The shaded area, dashed blue line, and solid red line, indicate the three fractional times to which the arcsine laws apply, namely $T_1$, $T_2$, and $T_3$, respectively.}
\end{figure}

Figure~\ref{fig2} shows a typical experimental trajectory of $\mathcal{I}_{n\tau}$ relative to the mean. Each step in this random walk was obtained by integrating the intensity up to the number of cycles indicated by the horizontal axis. Figure~\ref{fig2} also illustrates the definitions of $T_{1,2,3}$ using this particular trajectory.

\begin{figure}
    \centering
    \includegraphics{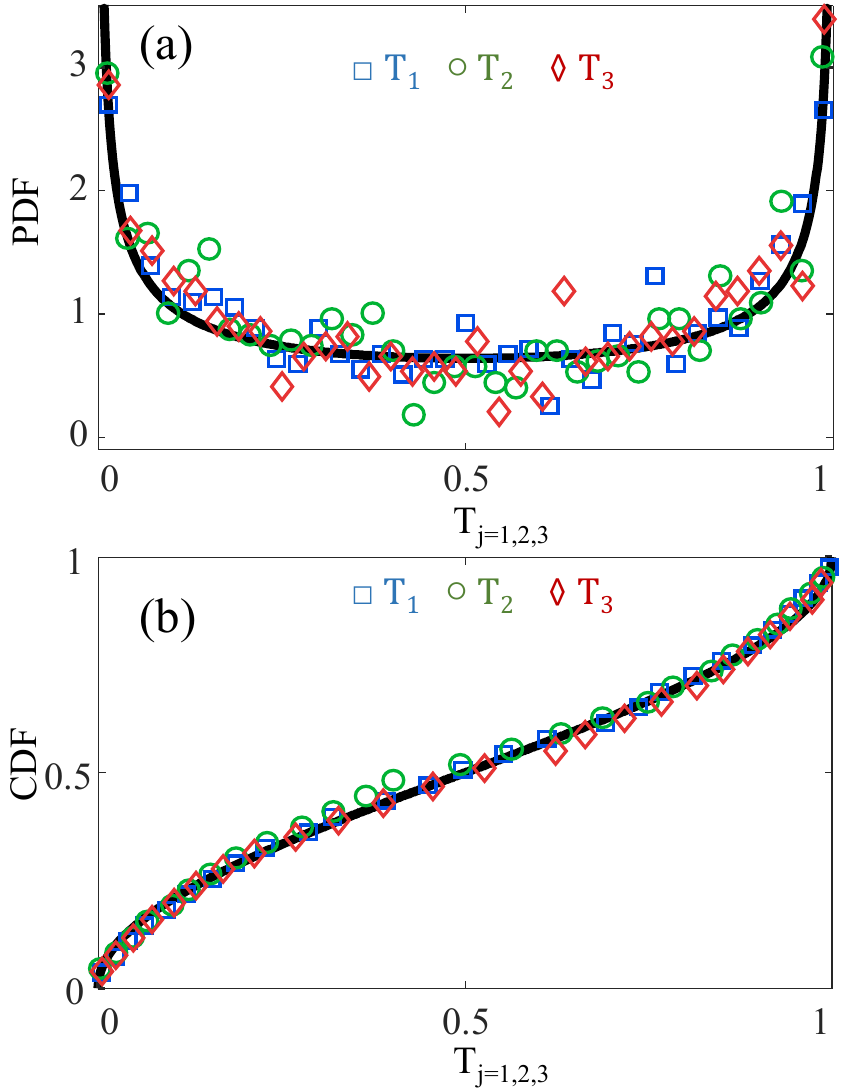}
    \caption{\label{fig3} (a) Probability distribution and (b) cumulative distribution for $T_{j=1,2,3}$ extracted from trajectories of the time-integrated intensity like the one shown in Fig.~\ref{fig2}.  Black curves in (a) and (b) are the analytical predictions of Equations~\ref{eq1a} and~\ref{eq1b}, respectively.}
\end{figure}

We analyzed $1000$ trajectories of $\mathcal{I}_{n\tau}$, each comprising $200$ cycles. The resultant probability distributions and cumulative distributions for $T_{j=1,2,3}$ are shown in Figures~\ref{fig3}(a) and ~\ref{fig3}(b), respectively. Blue squares, green circles, and red diamonds, correspond to $T_1$, $T_2$, and $T_3$, respectively. The black curves under the experimental data points are calculations of $\mathcal{P}(T_j)$ and $\mathcal{C}(T_j)$ using Equations~\ref{eq1a} and~\ref{eq1b}, respectively. The good agreement between theory and experiment demonstrates that the time-integrated transmitted intensity obeys the arcsine laws.

We also investigated the statistics of $T_{j=1,2,3}$ numerically, by solving Equation~\ref{eq2} using the xSDPE \textsc{Matlab} toolbox~\cite{xspde}.  We considered random walks of  $\mathcal{I}_{n\tau}$, and of the time-integrated field quadratures $\int_{0}^{n\tau} \alpha_R(t)$dt and $\int_{0}^{n\tau} \alpha_I(t)$dt. In supplemental material we present similar plots to those in Fig.~\ref{fig3}, evidencing that the arcsine laws are obeyed in all three cases~\cite{supp}. The time-integrated field quadratures are especially relevant to quantum-state tomography~\cite{Lvovsky}, and to sensing. In particular, Lau and Clerk showed that measuring a time-integrated field quadrature is optimal for linear sensing~\cite{Lau18}.

We furthermore investigated numerically the effects of a Kerr nonlinearity, by adding the term $U|\alpha|^2$ (with $U$ an effective photon-photon interaction strength) inside the parenthesis of Equation~\ref{eq2}. The Kerr-nonlinear term represents a non-conservative force which, for $U>0$, counteracts the detuning $\Delta$ in an intensity-dependent way. However, unlike $\Delta$, the Kerr-nonlinear term can lead to non-Gaussian statistics of $\alpha$. In particular, in supplemental material~\cite{supp} we show that the distribution of $\alpha$ in a Kerr-nonlinear cavity is squeezed and bimodal for large $A/\sqrt{\Gamma}$. Remarkably, the arcsine laws still hold in this case~\cite{supp}.  The importance of this result is linked to the many fundamental physics studies and technologies where time-integrated signals from Kerr-nonlinear resonators are analyzed. For example, Kerr-nonlinear resonators play a central role in recent studies of dissipative phase transitions~\cite{Carmichael15, Fink17, Casteels17, Rodriguez17, Fink18, Minganti18, Krimer19, Huybrechts20, Zhang21, Alaeian21, Soriente21}, symmetry breaking~\cite{Garbin22}, polariton blockade~\cite{Delteil19, Munoz19}, stochastic resonance~\cite{Abbaspour14, Braeckeveldt22}, non-reciprocity~\cite{Shi15, Sounas18, Cotrufo21_1, Cotrufo21_2}, and sensing~\cite{Heugel19, Rodriguez_PRAppl_2020, Peters22}.

\begin{figure}
    \centering
    \includegraphics{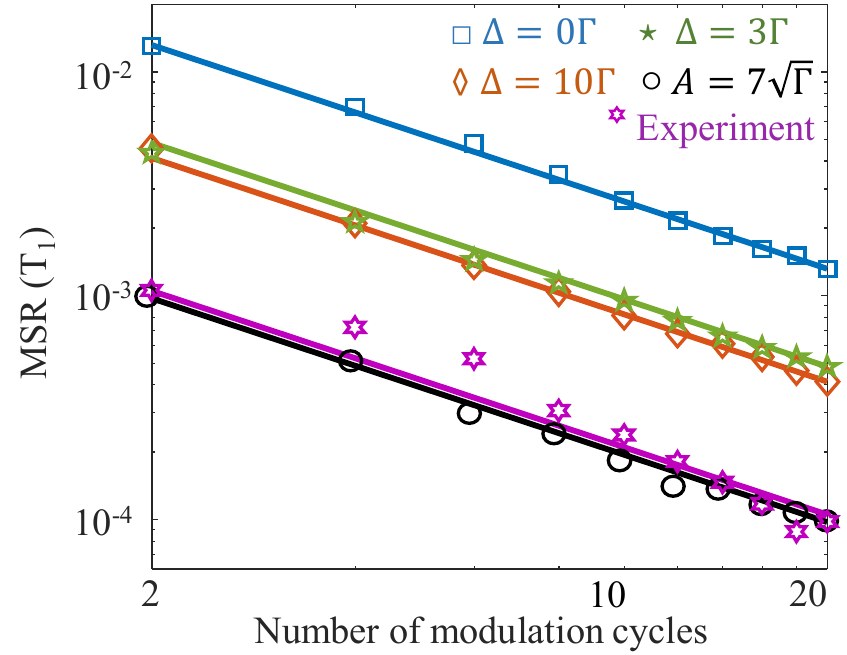}
    \caption{\label{fig4} Mean squared residuals between the arcsine distribution $\mathcal{C}(T_1)$ (Equation~\ref{eq1b}) and the finite time distribution $\mathcal{C}_{n \tau}(T_{j})$ as the transmitted intensity is integrated over an increasing number of cycles. Purple stars correspond to the experimental data in Figs.~\ref{fig1} and~\ref{fig3}. All other data is calculated numerically, using $10^4$ simulations with distinct noise realizations per data point. Each type of symbol corresponds to a different protocol, with different balance between conservative and non-conservative forces.  Blue squares, green stars, and orange diamonds, correspond to periodic modulations of the laser amplitude $A/\sqrt{\Gamma}$ between $5$ and $12$ while $\Delta/\Gamma$ is fixed to the value indicated in the legend.  Black circles correspond  to a periodic modulation of $\Delta/\Gamma$  between $-10$ and $10$ while  $A/\sqrt{\Gamma}$ is fixed to the value indicated in the legend. Solid lines of the same color as the data points are power laws with exponent $-1$ fitted to the data. $D=2 \sqrt{\Gamma}$ in all simulations; this choice does not alter the results. All results are for the fractional time $T_1$, but similar results hold for $T_2$ and $T_3$. All modulations have a period $\tau=50/ \Gamma$, but the results are the same for any $\tau \gg \Gamma^{-1}$.}
\end{figure}

Next we assess the convergence rate to the arcsine distribution as the number of integrated modulation cycles increases.  As indicators for covergence we use the mean square residuals (MSRs), given by $\frac{1}{n}\sum \left(\mathcal{C}(T_{j}) - \mathcal{C}_{n \tau}(T_{j})\right)^2$ with $\mathcal{C}_{n \tau}(T_{j})$ the finite time cumulative distribution. Figure~\ref{fig4} shows the MSRs for $T_1$, for various driving conditions. Similar results obtained for $T_{2}$ and $T_{3}$ are omitted for brevity. The purple stars in Fig.~\ref{fig4} are experimental data points obtained  by modulating the cavity length at fixed laser amplitude; this data corresponds to the results in Fig.~\ref{fig3}. The black circles are the corresponding numerical results, for constant laser amplitude $A=7 \sqrt{\Gamma}$ and periodically modulated $\Delta$.    Figure~\ref{fig4} also shows numerical results for 3 distinct constant $\Delta$ and periodically modulated $A$. These are the blue squares, orange diamonds, and green triangles.   By analyzing the MSRs for various driving conditions, we tested whether the balance between conservative and non-conservative forces influences the convergence rate to the arcsine distribution. Remarkably, the convergence rate is always the same. In the experiments and in all simulations, the MSRs decay with the total integration time by following a power law with exponent -1. This result holds for any modulation of $A$ at fixed $\Delta$, and for any modulation  of $\Delta$ at fixed $A$. We also analyzed numerically the more trivial case of constant $\Delta$ and $A$, which is a common driving condition in optical sensing. Again, we found the same universal convergence rate towards the arcsine distribution, regardless of the steady state properties.

To conclude, we demonstrated that the time-integrated intensity transmitted through a linear resonator obeys Levy's celebrated arcsine laws.  This asymptotic behavior of the cumulative distribution emerges at a universal rate with the integration time. Our results point to a new frontier of physics where optical resonators and extreme value statistics complement each other. On one hand, optical resonators are a powerful platform for probing emergent statistical structure under huge and precise variations of non-conservative forces driving the system out of equilibrium. The high optical frequencies and the availability of high-speed light detectors are ideally suited to characterize extreme deviations from the mean and/or rare events in stochastic dynamics across an unprecedented temporal range. On the other hand, by investigating ideas emerging from studies of extreme value statistics and stochastic thermodynamics, intriguing statistical properties of optical fields can be uncovered. Such is the case of the arcsine laws of light presented in this Letter. The discovery of emergent statistical structure in light can also impact many light-based technologies. For instance, time-integrated intensities and fields as studied in this paper are the quantities measured by every photodetector and thus every optical sensor. As such, understanding the effects of noise on these quantities can aid in the design of optical devices. Time-integrated intensities also play a prominent role in other coherently driven systems, such as superconducting circuits~\cite{Blais18}, magnon polaritons~\cite{Wang18}, and acoustic resonators~\cite{Fleury19}. Finally, an interesting perspective of our work is the study of emergent statistical structure in resonators with memory, such as a dye-filled~\cite{Klaers10, Alaeian17} or oil-filled~\cite{Geng20, Peters21} microcavities. This includes possible deviations from the arcsine laws which were derived for memoryless systems.

\section*{Acknowledgments}
\noindent This work is part of the research programme of the Netherlands Organisation for Scientific Research (NWO). We thank Jesse Slim, Age Tjalma, and Allard Mosk for a critical review of our manuscript, Aurelien Trichet for providing the concave mirror used in our experiments, and Christopher Jarzynski for stimulating discussions. We acknowledge funding from an ERC Starting Grant with project number 85269, and an ENW-XS grant with file number OCENW.XS21.1.110.



\providecommand{\noopsort}[1]{}\providecommand{\singleletter}[1]{#1}%

\end{document}